\documentclass[sigconf]{acmart}
\usepackage{booktabs} % For formal tables

\usepackage[T1]{fontenc}
\usepackage{epstopdf}
\usepackage{multirow}
\usepackage{balance}
\usepackage{amsfonts}
\usepackage{amsmath}
\usepackage{amssymb}
\usepackage[boxed, ruled,vlined,linesnumbered]{algorithm2e}
\usepackage{bbold}
\usepackage{algpseudocode}
\usepackage{microtype}
\usepackage{lmodern}
\usepackage{enumitem}

\usepackage{threeparttable}
\usepackage{bm}
\usepackage{multirow}
\usepackage{makecell}

\usepackage{subfig}
\usepackage{booktabs}
\usepackage{caption}

\usepackage{graphicx}

\newcommand{\nop}[1]{}
\newcommand{\eg}{e.g.}
\newcommand{\ie}{{\sl i.e.}}

\newtheorem{thm:def}{Definition}
\newtheorem{thm:eg}{Example}
\newtheorem{thm:lem}{Lemma}
\newtheorem{thm:obs}{Observation}

\newcommand{\svd}{\textsc{Truncated-SVD}\xspace}
\newcommand{\varimax}{\textsc{Varimax}\xspace}
\newcommand{\qassign}{\textsc{Query-Assign}\xspace}
\newcommand{\mixrate}{\texttt{mix\_rate}\xspace}

\newcommand{\dnn}{\mbox{\sf DPRM}}
\newcommand{\qcdnn}{\mbox{\sf QC-DPRM}}
\newcommand{\qcwdnn}{\mbox{\sf QC-WDPRM}}
\newcommand{\qcmtlnn}{\mbox{\sf QC-MTLRM}}

\setcopyright{rightsretained}

\begin{document}

% Copyright
\copyrightyear{2018} 
\acmYear{2018} 
\acmConference[CIKM '18]{The 27th ACM International Conference on Information and Knowledge Management}{October 22--26, 2018}{Torino, Italy}
\acmBooktitle{The 27th ACM International Conference on Information and Knowledge Management (CIKM '18), October 22--26, 2018, Torino, Italy}
%\acmDOI{10.1145/3269206.3272019}
%\acmISBN{978-1-4503-6014-2/18/10}

\title{Multi-Task Learning for Email Search Ranking \\ with Auxiliary Query Clustering}
\author{Jiaming Shen$^{1*}$, Maryam Karimzadehgan$^{2}$, Michael Bendersky$^{2}$, Zhen Qin$^{2}$, Donald Metzler$^{2}$}
\affiliation{%
  \institution{$^{1}$Department of Computer Science, University of Illinois Urbana-Champaign, IL, USA}
}
\affiliation{%
  \institution{$^{2}$Google Inc., Mountain View, CA, USA}
}
\affiliation{%
  \institution{$^{1}$js2@illinois.edu $\quad$ $^{2}$\{maryamk, bemike, zhenqin, metzler\}@google.com}
}

\thanks{*Work done while interning at Google.}

\fancyhead{}
\renewcommand{\shorttitle}{Multi-Task Learning for Email Search Ranking with Auxiliary Query Clustering}
\renewcommand{\shortauthors}{J. Shen et al.}

\begin{abstract}
	%!TEX root = main.tex
% UTF-8 encoding

User information needs vary significantly across different tasks, and therefore their queries will also differ considerably in their expressiveness and semantics. 
Many studies have been proposed to model such query diversity by obtaining query types and building query-dependent ranking models.
These studies typically require either a labeled query dataset or clicks from multiple users aggregated over the same document. 
These techniques, however, are not applicable when manual query labeling is not viable, and aggregated clicks are unavailable due to the private nature of the document collection, \emph{e.g.}, in email search scenarios.
In this paper, we study how to obtain query type in an unsupervised fashion and how to incorporate this information into query-dependent ranking models.
We first develop a hierarchical clustering algorithm based on truncated SVD and varimax rotation to obtain coarse-to-fine query types. 
Then, we study three query-dependent ranking models, including two neural models that leverage query type information as additional features, and one novel multi-task neural model that views query type as the label for the auxiliary query cluster prediction task. 
This multi-task model is trained to simultaneously rank documents and predict query types. 
Our experiments on tens of millions of real-world email search queries demonstrate that the proposed multi-task model can significantly outperform the baseline neural ranking models, which either do not incorporate query type information or just simply feed query type as an additional feature.

\end{abstract}

\begin{CCSXML}
<ccs2012>
<concept>
<concept_id>10002951.10003317.10003338</concept_id>
<concept_desc>Information systems~Retrieval models and ranking</concept_desc>
<concept_significance>500</concept_significance>
</concept>
</ccs2012>
\end{CCSXML}

\ccsdesc[500]{Information systems~Retrieval models and ranking}

\keywords{Email Search; Neural Ranking Model; Multi-Task Learning; Query Clustering}

\maketitle

%!TEX root = main.tex
% UTF-8 encoding
\section{Introduction}
User search queries come in very different, diverse flavors. For example, queries can be keywords, combinations of phrases, or just natural language sentences~\cite{Geng2008QueryDR}. Queries may also differ in length, grammatical structure, and ambiguity \cite{HangLi+al:2012}. Broder~\cite{broder2002taxonomy} showed that web search queries could be navigational, informational, or transactional, and different types of queries will prefer different ranking criteria.

Based on these observations, efforts have been made on obtaining different query types and building query-dependent ranking models. For example, Wen et al.~\cite{Wen2002QueryCU} proposed to cluster queries based on their clicked documents in search logs and treated each query cluster as the query type. Kang and Kim \cite{Kang2003QueryTC} classified queries into two categories using training data and then built two separate ranking models for two categories. Geng et al. \cite{Geng2008QueryDR} extracted $k$-nearest neighbors of each user-issued query in a labeled query pool and then constructed a ranking model based on this query subset. 
These methods have proven to be useful in the context of web search where either a labeled query set (for query classification) or click information aggregated across users is available. 
However, it is challenging to obtain/leverage such query type information or to build a query-dependent ranking model in a search scenario where manual query labeling is not viable due to privacy, and aggregated click data is unavailable. 

With web mail services offering larger and larger cloud storage, email search is quickly becoming an important research topic in information retrieval \cite{Carmel2015RankBT, Wang2016LearningTR, Bendersky2017LearningFU, Zamani2017SituationalCF, wang2018position}. 
In email search, users can only issue queries over their own private email collections. 
As shown in \cite{Carmel2015RankBT, Kim2017UnderstandingAM}, email search queries also exhibit a lot of diversity in form, as well as in intent. For example, for some queries like ``recent water bill'' and ``chase bank statement'', a reverse chronological ordering strategy would be preferable. On the other hand, for queries such as ``neural model papers'' and ``UMAI proposal'', relevance and content-based ranking would work better. As a result, some features (\emph{e.g.}, the recency of an email) will play a different role in the ranking model for different query types, and using a single universal ranking function is not suitable to model such query diversity. 

In this paper, we propose a query-dependent ranking framework that leverages auxiliary query clustering for email search ranking. 
To overcome the challenges that neither labeled queries nor aggregated clicks are available in email search, we apply an unsupervised hierarchical clustering algorithm based on truncated SVD \cite{Deerwester1990IndexingBL} and varimax rotation \cite{kaiser1958varimax} to obtain coarse-to-fine query types. 
The key idea is that queries of the same type will share the same or similar query attributes~\cite{Bendersky2017LearningFU}. 
Therefore, we can obtain query types by clustering queries based on these attributes. 
For example, we will cluster queries containing travel-related bigrams like ``Uber trip'' or ``Lyft itinerary'' and treat them as the same type. 

Once we have this query type information, intuitively, we can view each query's cluster as a feature and feed it into a ranking model. 
Based on this idea, we propose a pairwise neural ranking model. 
Given a query and a pair of documents with precisely one click, this model will directly learn the query/document representation and predicts the clicked document. 
As the query cluster is integrated in the query features, the model is trained to leverage such cluster information. 
The main drawback of this model is that it fails to distinguish between query cluster features and other query features. 
To deal with this problem, we propose the second model based on the wide and deep architecture \cite{Cheng2016WideD} and feed cluster-enhanced cross-product features into the wide part of network. 
In such a way, this model will treat the query cluster feature differently from other query features. 

These two models view the query type as an input feature, same as other query features. 
However, both pervious research \cite{Beutel2018LatentCM, Cheng2016WideD} and our own experiment results show that neural networks can be inefficient in modeling the interactions between features directly from input layers.
To solve this problem and to better leverage the query type information, we design our third ranking model using multi-task learning. 
This model views query cluster as the output label for auxiliary query cluster prediction task.
Specifically, the model adopts a multi-task objective function and is trained to simultaneously rank documents and predicts query clusters. 
In such a way, the query cluster information can be propagated to all intermediate layers of neural models and also helps the model to learn better query/document representations. 
Our experiment on tens of millions of real-world queries also demonstrates the effectiveness of this approach. 

In summary, this paper makes the following contributions:
\begin{enumerate}[leftmargin=*]
\item We develop a novel hierarchical clustering algorithm based on truncated SVD and varimax rotation. It works with high-dimensional input and returns multiple coarse-to-fine query types in a top-down fashion. 
\item We propose three query-dependent ranking models including a novel multi-task learning model that effectively leverages the query type information for improving the ranking performance. To the best of our knowledge, this is the first work that uses multi-task learning in the email search ranking.
\item We conduct a thorough evaluation of our proposed models using tens of millions of real-world email search queries. The experiment results show that our multi-task learning model can significantly outperform other baseline neural ranking models. 
\end{enumerate}
The rest of this paper is organized as follows. Section 2 discusses related work. Section 3.1 formalizes our problem and Section 3.2 presents our hierarchical query clustering algorithm. Then, we discuss three query-dependent ranking models In Section 3.3 and Section 3.4. In Section 4, we report and analyze the experimental results. Finally, we conclude the paper and discuss some future directions in Section 5.

%!TEX root = main.tex
% UTF-8 encoding
\section{Related Work}

\noindent \textbf{Email Search.}
With web mail sources offering larger and larger cloud storage, there is large amount of emails stored in the cloud. This large amount of data calls for effective email search capabilities. 
Compared with web search, email search has two unique challenges. First, we cannot obtain explicit relevance judgments in the form of labeled $\langle$query, document$\rangle$ pairs due to the privacy issue.. To solve this problem, we use the click data as implicit relevance judgments. Second, we cannot directly leverage the ``wisdom of crowds'' to aggregate clicks on the same document across different users. Consequently, many learning methods based on co-click data in web search are no longer applicable. To deal with this problem, Bendersky et al. \cite{Bendersky2017LearningFU} proposed an attribution parameterization method that uses click-through rate of query and document attributes (\emph{e.g.}, frequent $n$-grams) to learn a ranking model. Zamani et al. \cite{Zamani2017SituationalCF} developed a context-aware wide and deep network model for semantic matching. In this paper, we propose a neural model that can leverage query cluster information and show the new model can significantly outperform previous approaches. 

\vspace{-0.1em}
\begin{figure*}[!t]
\center
\includegraphics[width = 0.8\textwidth]{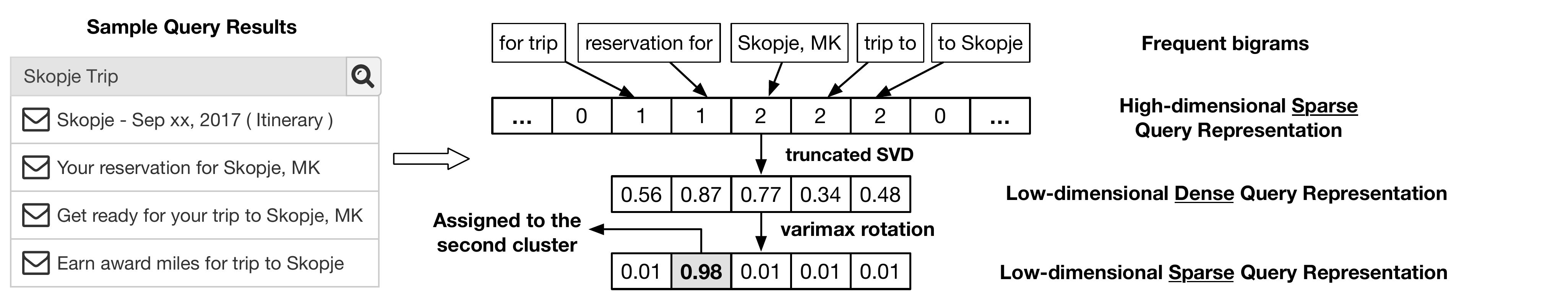}
\vspace{-0.3cm}
\caption{An illustrative example showing the query raw representation and multiple low-dimensional representations learned during query clustering process.}
\label{fig:queryClustering}
\vspace{-0.3cm}
\end{figure*}

\smallskip
\noindent \textbf{Query-dependent Ranking Models.}
Since user's information need can vary across different tasks, search queries can be grouped into different types.
Broder \cite{broder2002taxonomy} showed that web search queries can be navigational, informational, or transactional. Similarly, Carmel et al. \cite{Carmel2015RankBT} showed that in email search, for some queries users prefer returned emails sorted by time, while for other queries they prefer returned emails ranked by textual relevance. Intuitively, we would like to leverage such query type information and exploit different ranking models for different query types. 

To build such query-dependent ranking models, we need to obtain the query type first. Many efforts have been made on query classification \cite{Shen2006BuildingBF, Beitzel2007VaryingAT, Cao2009ContextawareQC}, which require a labeled set of queries for each query type. Such query labeling is costly, and thus some studies proposed to obtain query types by unsupervised query clustering instead of supervised query classification. Wen et al. \cite{Wen2002QueryCU} used the DBSCAN algorithm to cluster queries based on user logs. Sadikov et al. \cite{Sadikov2010ClusteringQR} applied a probabilistic model to cluster query refinements based on document co-click and session co-occurrence. Geng et al. \cite{Geng2008QueryDR} proposed to extract $k$-nearest neighbors of the user issued query in a labeled query pool first and then to construct a ranking model based on this query subset. All these methods require either labeled queries or cross-user co-click data, which is difficult or even impossible to obtain in the email search setting. In this paper, we propose a completely unsupervised hierarchical clustering method and seamlessly incorporate query clusters of different granularities in a neural ranking model. 

\smallskip
\noindent \textbf{Neural Ranking Models.}
Classical learning to rank models~\cite{Burges2005LearningTR, Burges2010FromRT} typically require hand-crafted features as input. Neural models, on the other hand, attempt to directly learn query/document representations and avoid tedious feature engineering. 
Salakhutdinov and Hinton \cite{Salakhutdinov2009SemanticH} proposed to adopt auto-encoder architecture which learns binary representations of documents. More recently, Huang et al. \cite{Huang2013LearningDS} proposed DSSM architecture that represents the query and document as a bag-of-character-trigrams and trains on clickthrough data. As a special type of Siamese architecture \cite{Bromley1993SignatureVU}, DSSM model uses a fully connected layer for query and document models. More sophisticated architectures are also explored, including those using convolutional layers \cite{Gao2014ModelingIW, Shen2014LearningSR}, recurrent layers \cite{Palangi2014SemanticMW, Palangi2016DeepSE}, and recursive layers \cite{Tai2015ImprovedSR}. Besides those representation-based models, other interaction-based models are proposed. Lu and Li \cite{Lu2013ADA} proposed DeepMatch which first compares different parts of the query with different parts of the document and then uses a feed forward network for computing the relevance score. Hu et al. \cite{Hu2014ConvolutionalNN} used stacked convolutional neural network models for matching two sentences. Other interaction-based models include \cite{Mitra2017LearningTM, Pang2016ASO}. For a complete literature review on neural ranking models, please refer to \cite{Mitra2017NeuralMF}. 
%!TEX root = main.tex
% UTF-8 encoding
\section{Methodology}

We first formulate our problem and define notations. 
Then we describe our hierarchical query clustering algorithm. 
Finally, we present three query-dependent ranking models, including one novel multi-task neural ranking model that is trained to rank documents and predict query types, simultaneously.

%% 3.1
\vspace{-0.5em}
\subsection{Problem Formulation}

Let $\mathbb{Q} = \{Q_{1}, \dots, Q_{M} \}$ denote a set of $M$ queries.
Each $Q_{i} = (\mathbf{q_{i}}, [ (\mathbf{d_{i1}}, c_{i1}), (\mathbf{d_{i2}}, c_{i2}), \dots, (\mathbf{d_{iN}}, c_{iN}) ] )$ represents the $i^{th}$ query in the query set that contains the query feature $\mathbf{q_{i}}$ and the document feature $\mathbf{d_{ij}}$ along with its click information $c_{ij}$ for each document. If the $j^{th}$ document is clicked for the $i^{th}$ query, we set $c_{ij}=1$, otherwise, we set $c_{ij}=0$.
Given the training query set $\mathbb{Q}$, we learn a model to find a scoring function $f(\cdot)$ that can minimize the empirical loss defined as:
\vspace{-0.1em}
\scriptsize
\begin{displaymath}
L_{\mathbb{Q}}\left(f\right) = \frac{1}{|\mathbb{Q}|} \sum_{Q_{i} \in \mathbb{Q}} l(Q_{i}, f), 
\end{displaymath}
\normalsize
where $l(Q_{i}, f)$ denotes the loss of function $f$ on query $Q_{i}$. 

In this paper, we use the following pairwise loss function based on the pairwise topology described in \cite{dehghani2017neural}, where it was found to be useful for weakly supervised learning:
\vspace{-0.1em}
\scriptsize
\begin{eqnarray*}
l(Q_{i}, f) & = & \sum_{ \langle \mathbf{d_{ia}}, \mathbf{d_{ib}} \rangle \in Q_{i}} -y_{ab}\log(p_{ab}) - (1-y_{ab})\log(1-p_{ab}), \\
p_{ab} & = & f(\langle \mathbf{q_{i}}, \mathbf{d_{ia}}, \mathbf{d_{ib}} \rangle), 
\end{eqnarray*}
\normalsize
where $y_{ab}$ equals to 1 if $\mathbf{d_{ia}}$ is more relevant than $\mathbf{d_{ib}}$ (namely, $c_{ia} = 1$ and $c_{ib} = 0$) and equals to 0, otherwise. 

Given a new query $Q_{t} = (\mathbf{q_{t}}, [\mathbf{d_{t1}}, \mathbf{d_{t2}}, \dots, \mathbf{d_{tN}}])$, we use the learned $f(\cdot)$ to score each document in $Q_{t}$ as follows:
\vspace{-0.1em}
\scriptsize
\begin{displaymath}
score(\mathbf{d_{ta}}) = \frac{1}{N} \sum_{\mathbf{d_{tb}} \neq \mathbf{d_{ta}}} f(\langle \mathbf{q_{t}}, \mathbf{d_{ta}}, \mathbf{d_{tb}} \rangle).
\end{displaymath}
\normalsize
After we obtain the score for each document, we can either directly rank documents based on these scores or feed them as additional ranking signals into a learning-to-rank framework. 

%% 3.2
\vspace{-0.5em}
\subsection{Hierarchical Query Clustering}

In this section, we describe how we represent the email search query and how we conduct query clustering based on the query representation. 

\smallskip
\noindent \textbf{Query Representation.}
Due to privacy issues in email search, we can only access very limited query features that do not reveal any information about the user, such as frequent $n$-grams\footnote{All available feature details are presented in the experiment section.}. 
Therefore, we follow the approach in \cite{Geng2008QueryDR} and first apply a reasonable baseline ranker (\emph{e.g.}, BM25) to obtain a pre-ranked document list. Then, we aggregate the features of top-ranked documents in this pre-ranked list and use them to enhance the query representation. The philosophy is that although these documents may not be in the optimal ranking order, they can still provide useful information for a query. A similar idea is proved useful in \cite{dehghani2017neural}. 

We explain the details by following an example in Figure \ref{fig:queryClustering}. Suppose a user issues a query ``Skopje trip'' and obtains a preliminarily ranked email list. Based on the top 4 emails in this ranked list, as well as the query itself, we can collect five frequent 2-grams, one from the query and the other four from documents. Then we represent this query using a high-dimensional sparse vector. For example, the bigram ``to Skopje'' appears twice in the top 4 ranked emails and thus the feature value corresponding to this bigram equals to 2. 

\smallskip
\noindent \textbf{Query Clustering.}
As the generated query feature vector is of high dimensionality but very sparse, we need to design an algorithm which can work with high-dimensional input and can leverage such sparsity.
Also, we want our algorithm to scale to billions of examples with potentially thousands of clusters. 
Traditional clustering algorithms such as $k$-means and DBSCAN can not satisfy all these requirements at the same time \cite{han2011data}, and therefore, we design a hierarchical clustering algorithm which returns multiple coarse-to-fine query clusters by learning low-dimensional query representations.

At the high level, query clusters are constructed recursively in a top-down manner. Initially, all queries are put in the root, denoted as the level-0 cluster. Then, to assign each query to one of the level-1 clusters, we do the following two steps. First, we train a truncated SVD model \cite{Deerwester1990IndexingBL} based on the query high-dimensional sparse representation. We choose to use truncated SVD model because it can leverage the sparsity of input query feature and outputs a dense low-dimensional representation of this query, as shown in Figure \ref{fig:queryClustering}. Second, we apply a varimax rotation \cite{kaiser1958varimax} on top of the dense low-dimensional query representation. The varimax rotation tries to project each query vector into as few axes as possible which returns a sparse low-dimensional representation of the query. Each dimension in the final sparse low-dimensional query vector represents a level-1 cluster and the dimension with the highest score is taken as the query's level-1 cluster assignment. For example, the sample query in Figure \ref{fig:queryClustering} is put in the second level-1 cluster as it has the highest score in the second dimension. 

After all queries are put into one of the level-1 clusters (\emph{e.g.}, cluster-1, cluster-2, ...), we proceed to construct all level-2 clusters (\emph{e.g.}, cluster-1.1, cluster-1.2, cluster-2.1, cluster-2.2, ...) under each level-1 clusters. To achieve this, we learn a new truncated SVD model based on all queries that reside in one specific level-1 cluster and apply the varimax rotation again. We essentially re-learn a new sparse low-dimensional query representation because we want to take advantage of the fact that different subsets of the data may be best described by their own distance metrics. Also, note that we can construct the level-2 clusters under one specific level-1 cluster (\emph{e.g.}, cluster-1) without any knowledge on other level-1 clusters (\emph{e.g.}, cluster-2, cluster-3, ...). Therefore, we can implement this hierarchical clustering algorithm in a distributed fashion and allow it to scale to billions of examples. 

Algorithm 1 summarizes our divisive hierarchical query clustering algorithm (non-recursive version). Given a collection of queries $\mathbb{Q}$, the depth of hierarchy $D$ and the number of branches in each level $B$, it first puts all queries in the root cluster node. Then, for each selected cluster $c$ of depth $d$, if the number of queries in it passes the minimal threshold (line 7), the algorithm will fit the subspace based on all the queries in $c$ and then learn their sparse low-dimensional representations $F_{c}$. Otherwise, this leaf cluster $c$ will be pruned. After that, it constructs $B$ sub-clusters and assigns each query $q_{i} \in S_{c}$ to one of them based on $F_{c}$. 
In such a way, each query will be assigned to at most one cluster per hierarchy level. 

%% Algorithm
\small
\begin{algorithm}[!t]
  \label{alg:clustering}
  \caption{Divisive Hierarchical Query Clustering}
  \KwIn{A collection of queries $\mathbb{Q}$; the depth of hierarchy $D$; the number of branch in each level $B$; the minimal required examples in each leaf node $E$.}
  \KwOut{A hierarchical cluster tree $T$ of queries $\mathbb{Q}$.}
  Assign all queries to root cluster at depth $d=0$\;
  Initialize $T \gets \emptyset$\;
  \For{depth $d$ ~ from 0 to $D$ }  { 
  	\For{cluster $c$ at depth $d$} {
		$S_{c} \gets$ queries assigned to $c$\;
		\If{$d = D$ } { 
			\If {$|S_{c}| < E$} {
				$T \gets T - \{c\}$ // Prune this leaf node\;
			}
			Continue\;
		}
		$F_{c} \gets \varimax(\svd(S_{c}, B))$\;
		$N_{1}, \ldots, N_{B} \gets \qassign(F_{c})$\;
		\For {sub cluster index $i$ ~ from 1 to $B$} {
			$T \gets T \cup N_{i}$\;
		}
	}
  }
  Return hierarchical cluster tree $T$\;
\end{algorithm}
\normalsize
\vspace{-0.5em}

%% Figure: all models
\begin{figure*}[!tbp]
	\centerline{\includegraphics[width=1.00\textwidth]{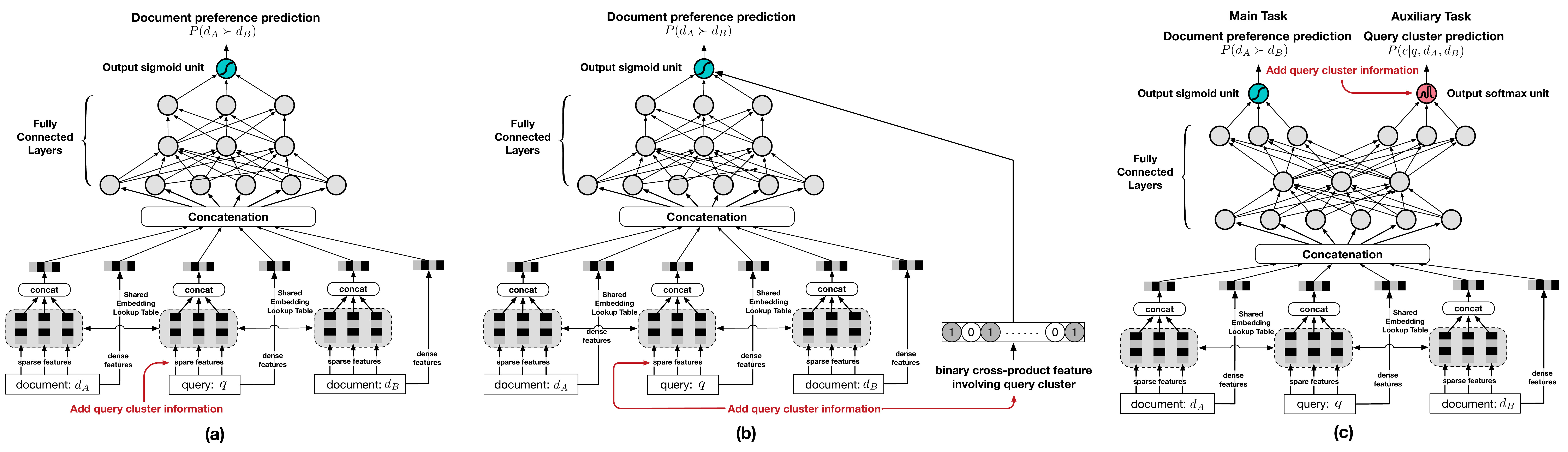}}
	\vspace{-0.2cm}
	\caption{The network architecture of (a) \qcdnn. (b) \qcwdnn. (c) \qcmtlnn. Note that the query cluster information in \qcdnn~and \qcwdnn~contributes in a ``bottom-up'' fashion from the input layer, while the \qcmtlnn~models it as the label for auxiliary task thus enables it to contribute in a ``top-down'' fashion.}
	\label{fig:models-all}
	\vspace{-0.2cm}
\end{figure*}

%% 3.3 Pairwise query-dependent ranking models
\subsection{Pairwise query-dependent ranking models}

We describe two pairwise neural ranking models in this section. 
The first one is motivated by the semantic matching model in \cite{Huang2013LearningDS}, and especially its application in the email search setting \cite{Zamani2017SituationalCF}. 
Instead of using the \emph{pointwise} topology in \cite{Huang2013LearningDS}, we adopt a \emph{pairwise} topology which is proved to be effective in the weakly supervised setting \cite{dehghani2017neural}. 
Following we refer to such \underline{D}eep \underline{P}airwise \underline{R}anking \underline{M}odel as \dnn.

The network architecture of \dnn~is shown in Figure \ref{fig:models-all}(a). 
The input includes a query $q$ and a pair of documents associated with this query. 
Specifically, one document is clicked while the other one is not clicked. 
The raw input feature of query and documents are described in Section~\ref{subsec:data_and_features}. 
Notice that some features are sparse categorical features (\eg, frequent subject $n$-grams) while the others are dense continuous features (\eg, email's recency). 
We feed the sparse features (for both query and documents) into an embedding layer which directly learns their dense representations\footnote{
We will replace all sparse features with frequency less than a threshold (in the training corpus) with a special unknown token UNK. 
Then, during the testing time, an unseen sparse feature will also be treated as UNK and thus has the same embedding of UNK.}.

After obtaining the query/document representations, we concatenate them and feed them into a stack of hidden layers. 
These hidden layers are fully connected, and the representation in $(i+1)^{th}$ hidden layer is obtained by mapping the representation in $i^{th}$ hidden layer as follows:
\small
\begin{displaymath}
\mathbf{h_{i+1}} = g(\mathbf{W_{i}^{T}} \cdot \mathbf{h_{i}} + \mathbf{b_{i}}),
\end{displaymath}
\normalsize
where $\mathbf{W_{i}}$ is the transformation matrix from layer $i$ to layer $i+1$, the $\mathbf{b_{i}}$ is the bias vector, and the activation function $g(\mathbf{x}) = \max(\mathbf{x}, 0)$ is the rectifier linear unit (ReLU) applied on each dimension of $\mathbf{x}$. 
Finally, in the output layer, we use one fully connected neuron with sigmoid activation function $\sigma(z) = \frac{1}{1+e^{-z}}$ to calculate the probability that document $d_{A}$ is preferred (i.e., more likely to be clicked) than document $d_{B}$. 
To incorporate query cluster information in \dnn, we treat each query's clusters as additional sparse features and feed them into the embedding layer. Similarly, we refer to such \underline{Q}uery \underline{C}luster aware \dnn~as \qcdnn.

One limitation of \qcdnn~is that it models the sparse query cluster feature and other sparse query features (\eg, query language) in the same manner. 
To explicitly differentiate these two types of features, we propose the second model based on the wide \& deep model similar to the one in \cite{Cheng2016WideD}. 
Our model architecture is shown in Figure \ref{fig:models-all}(b). 
The key idea is to jointly train a wide linear model (with potential cross-product features) as well as a deep neural model, and therefore combine the benefits of memorization and generalization. 
This simple approach is proved useful for building recommender system \cite{Cheng2016WideD}, and modeling query context features \cite{Zamani2017SituationalCF}.

In the context of email search ranking, we manually design a set of binary cross-product features involving query clusters and feed them into \emph{the wide part of the model}. 
These cross-product features can capture important feature interactions and add non-linearity to the wide linear model. 
For example, a cross-product feature ``\textsf{query\_cluster=1 AND language=English}'' is activated (\ie, equals to 1) if and only if this query is in cluster-1 and is of language English.
By requiring each cross-product feature in the wide part of model to contain one query cluster feature, we enable the model to differentiate between the query cluster feature and other sparse query features. 
We refer to such model as \qcwdnn~in the remainder of this paper. 

%% 3.4
\vspace{-0.4em}
\subsection{Multi-task query-dependent ranking model}
The \qcdnn~model and \qcwdnn~model essentially treat the query cluster information as an input feature.
However, both previous research \cite{Beutel2018LatentCM, Cheng2016WideD} and our own experiment results show that neural networks can be inefficient in modeling the interactions between features directly from input layers.
To solve this problem and better leverage the query type information, we take an alternative approach in this section and treat the query cluster as the ``label'' of an auxiliary query cluster prediction task.
We propose our third ranking model based on the multi-task learning framework.
Given a query and a pair of documents, this model simultaneously predicts which document is more likely to be clicked and which cluster(s) this query will reside in. 
The query cluster information will be propagated to every intermediate layer in a top-down fashion, which allows our model to learn a better representation of query and documents. 
As shown in \cite{Liu2015RepresentationLU}, good query/document representations learned by incorporating more knowledge from different tasks can significantly improve the ranking performance. 

The architecture of our proposed query-dependent multi-task ranking model is shown in Figure \ref{fig:models-all}(c). The lower layers are shared across two tasks. The top layers are different in each task. For the task of document preference prediction, we use fully connected hidden layers and an output layer composed of one single neuron. The output layer returns the probability that document A is preferred to document B. Given a query $q$, a pair of documents $\langle d_{A}, d_{B} \rangle$ with precisely one clicked, we define the following log-loss for this query:
\small
\begin{eqnarray}
l^{rank}(q) & = & -y_{ab}\log(p_{ab}) - (1-y_{ab})\log(1-p_{ab}), \\
p_{ab} & = & P(d_{A} \succ d_{B}),
\end{eqnarray}
\normalsize
where $y_{ab}$ equals to 1 if document $d_{A}$ is clicked and equals to 0 otherwise.

For the task of predicting query clusters, we use one additional fully connected hidden layer and a softmax output layer, on top of the shared layers. The dimensionality of the final softmax layer is the same as the number of query clusters, and it outputs the query cluster distribution. We use the cross-entropy loss as the objective for this task:

\small
\vspace{-0.5em}
\begin{eqnarray}
l^{cluster}(q) & = & -\sum_{c \in C} p_{c} \log \hat{p_{c}}, \\
\hat{p_{c}} & = & p(c | q, d_{A}, d_{B}), 
\end{eqnarray}
\normalsize
where $p_{c}$ is the true distribution on query cluster space\footnote{As we are using a hierarchical clustering algorithm, each query may reside in multiple query clusters (\emph{e.g.}, cluster-1, cluster-1.1, cluster-1.1.2). Suppose we have 100 valid query clusters (of different granularities), then this is essentially a distribution over these 100 query clusters (with only three non-zero values).}. 

In order to combine these two tasks, we define a joint multi-task objective function, as follows:
\small
\vspace{-0.2em}
\begin{equation}
L(\Theta) = \frac{1}{|Q|}  \sum_{q \in Q} \left( l^{rank}(q) + \lambda \cdot l^{cluster}(q) \right),
\end{equation}
\normalsize
where $\Theta$ represents all the neural network parameters and $\lambda$, named \mixrate, is used to balance the ranking loss and the query cluster prediction loss. We study how $\lambda$ influences the model performance in Section \ref{sec:query-specific-eval}.

Note that the above objective function can be transformed into the following form:
\small
\vspace{-0.2em}
\begin{equation}
L(\Theta) = \frac{1}{|Q|}  \sum_{q \in Q}  l^{rank}(q) + \lambda \left(  \frac{1}{|Q|}  \sum_{q \in Q} l^{cluster}(q) \right).
\end{equation}
\normalsize
The first term on the right-hand side of the equation is essentially the objective function used in \qcdnn~and \qcwdnn. The second term derived originally from the loss on query cluster prediction can be seen as a regularization (as it will influence how feature representations are learned) and is controlled by the parameter $\lambda$. In the remainder of this paper, we refer this multi-task ranking model as \qcmtlnn. 
%!TEX root = main.tex
% UTF-8 encoding
\vspace{-0.1cm}
\section{Experiments}

%% Table 1: describing features
\small
\begin{table*}[!tbp]
\centering
\caption{\small Summary of the query and document features used for clustering queries and learning ranking models.}
\vspace{-0.3cm}
\scalebox{0.9}{
\begin{tabular}{|l|l|c|}
\hline
\textbf{Feature Type} & \textbf{Descriptions} & \textbf{Usage} \\
\hline
Content & List of frequent $n$-grams appearing in the query text and the email subject & Cluster queries \\
	      & \emph{e.g., ``Class schedule on Friday morning'' $\rightarrow$ [``class schedule'', ``Friday morning'']}. & Learn ranking models \\
\hline
Category & Small set of commonly used email labels & Cluster queries \\
	       & \emph{e.g., Promotions, Forums, Purchases, and etc.} (see \cite{Bendersky2017LearningFU} for detailed label examples) & Learn ranking models \\
\hline
Structure & Frequent machine-generated email subject templates & \multirow{3}{*}{\makecell{Cluster queries\\Learn ranking models}}  \\
		& \emph{e.g., Your trip confirmation number 12345  $\rightarrow$ Your trip confirmation number *} &  \\
		& (see \cite{Bendersky2017LearningFU} Table 2 for more details on structure features) &  \\
\hline
Situational & Temporal and Geographical features of current search request & \multirow{2}{*}{\makecell{Learn ranking models}} \\
		& \emph{e.g., Friday, 8:00pm, USA, Japan} (see \cite{Zamani2017SituationalCF} for more details) &  \\
\hline
\end{tabular}}
\label{tab:features}
\vspace{-0.15cm}
\end{table*}
\normalsize

In this section, we first describe the data sources and features used in our experiments. 
Then, we evaluate our proposed models and study the influence of some important model hyper-parameters.
Finally, we feed the output of each model as an additional signal into an end-to-end ranking pipeline, and explore how much improvement can be achieved.

\vspace{-0.5ex}
\subsection{Data}\label{subsec:data_and_features}
To the best of our knowledge, there is no publicly available \emph{large-scale} email search dataset for training neural ranking models, probably because it is too private and too sensitive. 
Therefore, in this paper, we use the data derived from the search click logs of a commercial email service. 
The training set contains approximately 66 million queries, and there are about 4 million and 9 million queries in the development and testing set, respectively. 
All queries in the development set are issued strictly later than all queries in the training set, and all queries in the testing set are issued strictly later than all queries in the development set\footnote{Namely, we will not train a model using queries issued in May 2018 and test its performance on queries issued in April 2018.}. 
We construct datasets in such a way to avoid the potential data leakage problem. 
Each query has six candidate documents with precisely one clicked\footnote{These six candidates are presented in the dropdown menu while users type their queries in the search bar but before they click the search bottom. When users find the target email, the system will direct them to the exact email and thus generates precisely one click.}, and the purpose is to rank these six documents for the given query. 
Such large scale dataset also shows the scalability of our proposed models. 

The original datasets are anonymized based on $k$-anonymity approach \cite{Sweeney2002kAnonymityAM}, and thus the only content features we can access are the query and document $n$-grams that are frequent in the whole corpus. We use those frequent $n$-grams, as well as some other category and structure features \cite{Bendersky2017LearningFU} for query clustering. Besides these features, we also use the situational features including the language and timestamp of search requests to learn the ranking models, as described in \cite{Zamani2017SituationalCF}. All features are summarized in Table \ref{tab:features}. 

We use click data as ground truth labels to learn and evaluate our proposed ranking models, which is a standard practice for email search \cite{Carmel2015RankBT, Wang2016LearningTR, Bendersky2017LearningFU, Zamani2017SituationalCF}. Furthermore, we apply the position bias correction techniques \cite{Craswell2008AnEC, Wang2016LearningTR, wang2018position} to reduce the noise in click data during our model training.

%%% Section 4.2
\subsection{Query-dependent Ranking Model Evaluation}
\label{sec:query-specific-eval}
\vspace{-0.1em}
\subsubsection{Experimental Setup}
\hfill

We implement our neural network models using TensorFlow. We set the depth of hierarchy $D$ to be 3 and the number of branches $B$ to be 7. We analyze the effect of these clustering hyper-parameters in Section 4.2.3. We tune the neural network hyper-parameters over the development set as follows: we sweep the learning rate between \{0.05, 0.08, 0.1, 0.15, 0.3\}, the dropout rate between \{0.2, 0.3, 0.4, 0.5, 0.6, 0.7, 0.8, 0.9\}, the number of hidden layers between \{3, 4\}, the size of first hidden layers between \{64, 128, 256, 512, 768, 1024\}, the embedding dimension between \{20, 30, 40, 50, 100\}, and the optimization algorithm between \{Adagrad \cite{Duchi2010AdaptiveSM}, Adam \cite{kingma2014adam}\}. For each model, we select the combination of hyper-parameters that have the best performance on the development set and report its result on the testing set. 

\vspace{-0.1cm}
\subsubsection{Evaluation Metric}
\hfill

We evaluate model performance using mean reciprocal rank (MRR). Since each query has precisely one clicked document, the MRR is calculated as follows:
\small
\vspace{-0.1em}
\begin{displaymath}
MRR = \frac{1}{|Q|}  \sum_{i=1}^{|Q|} \frac{1}{rank_{i}},
\end{displaymath}
\normalsize
where $Q$ denotes the evaluation set and $rank_{i}$ represents the rank position of the clicked document for $i^{th}$ query in the evaluation set.

Similar to \cite{Zamani2017SituationalCF}, we also use success@1 and success@5 to evaluate our models. The success@$k$ measures the percentage of queries for which the clicked message is ranked within the top-$k$ results\footnote{In fact, success@1 is the precision of document ranked at the first position, and success@5 is the accuracy of the system in \emph{not} retrieving the correct (clicked) document at the last position.}\cite{Carmel2015RankBT}.
Finally, to compare the performance of multiple models, we conduct statistical significance test using the two-tailed paired $t$-test with 99\% confidence level. 

\vspace{-0.1em}
\subsubsection{Results and Discussion}
\hfill

In this section, we first show some query $n$-gram clusters. Then, we empirically study whether and how the query cluster feature can help improve the ranking results. Finally, we analyze how some important hyper-parameters affect the model performance. 

\begin{figure*}[!t]
\centering
    \subfloat[\small \label{fig:query_cluster_features}]{%
      \includegraphics[width=0.3\textwidth]{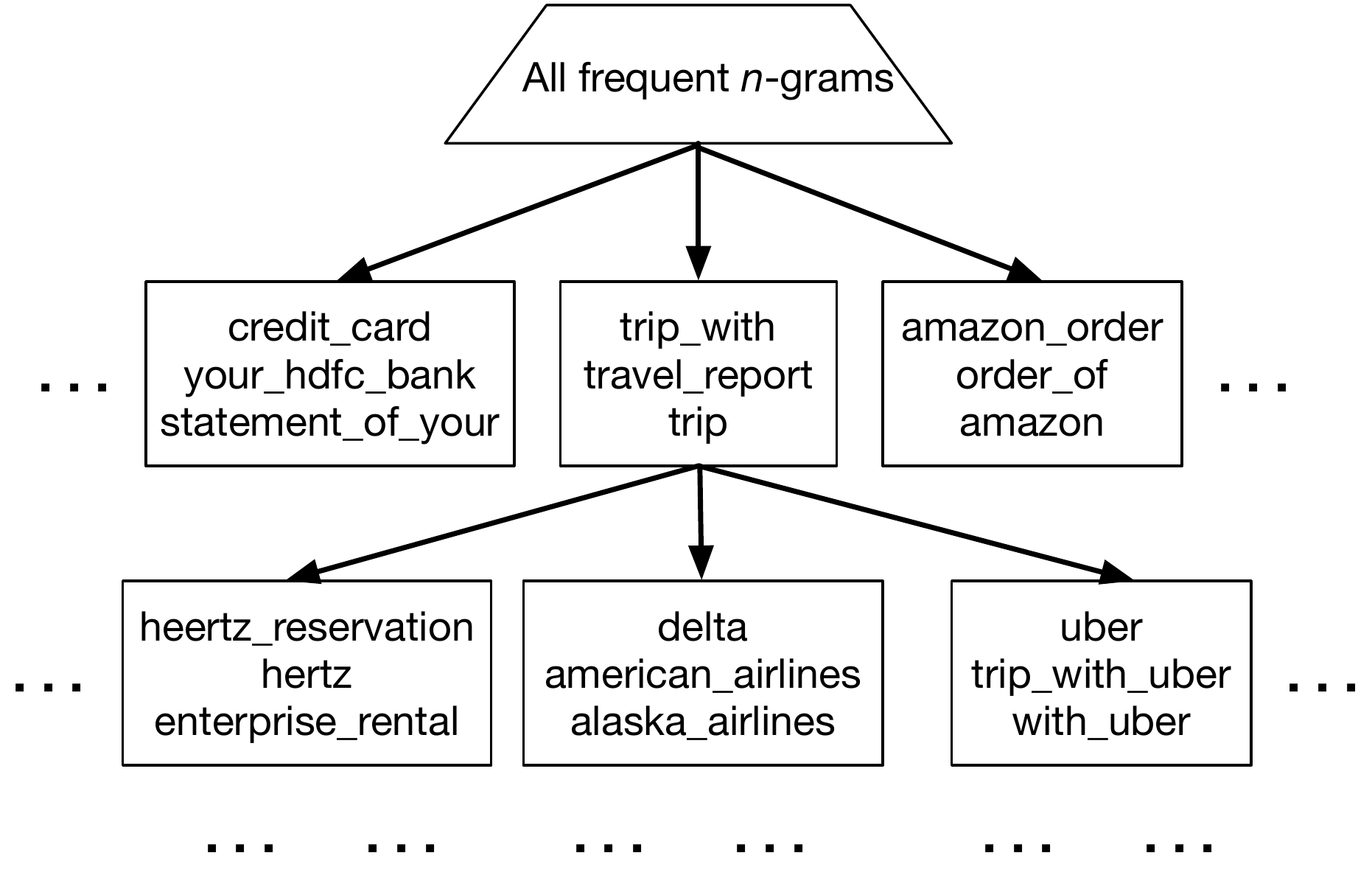}
    }
    \subfloat[\small \label{fig:mixRateAnalysis}]{%
      \includegraphics[width=0.3\textwidth]{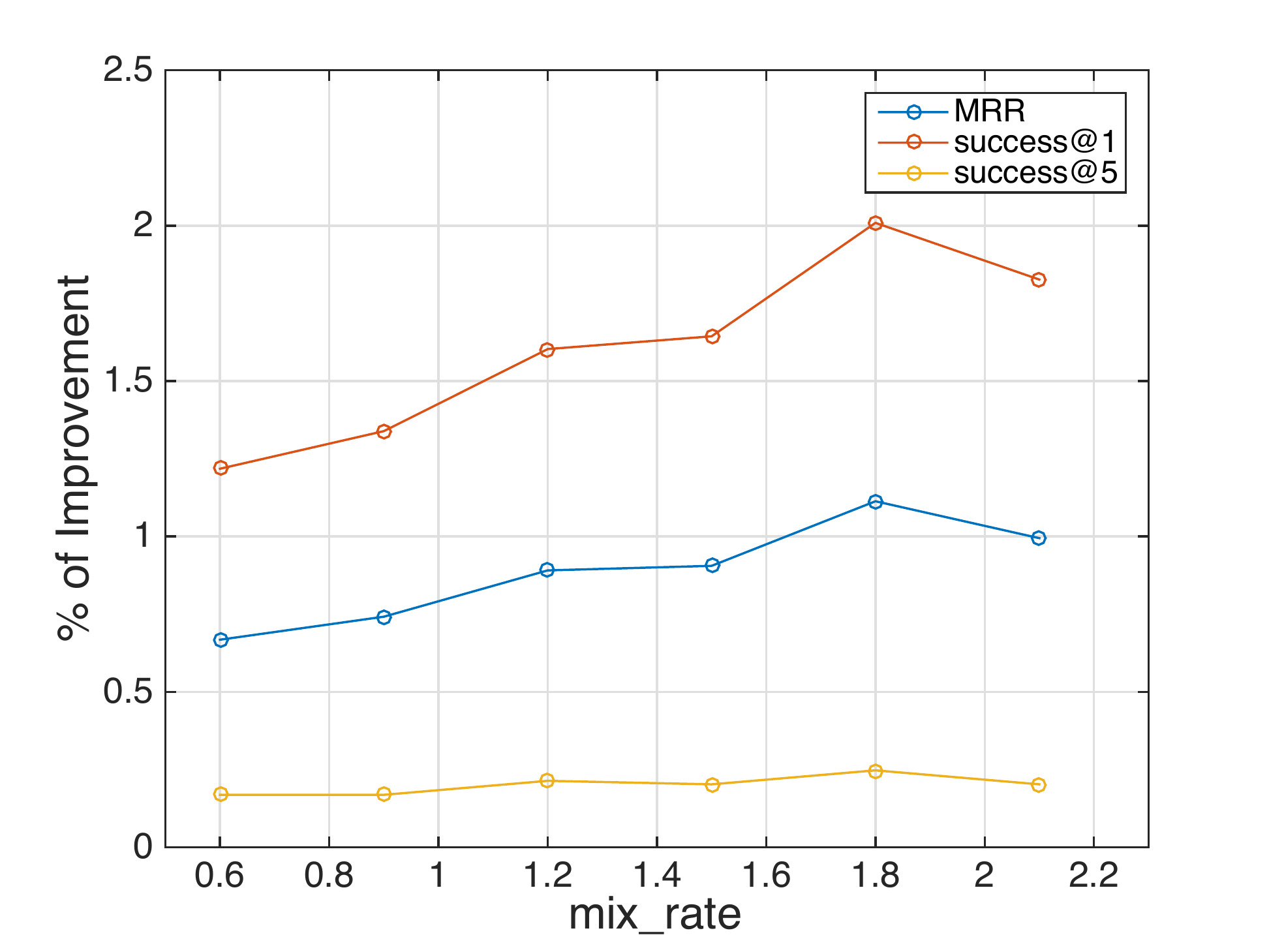}
    }
    \subfloat[\small \label{fig:clusterNumberAnalysis}]{%
      \includegraphics[width=0.3\textwidth]{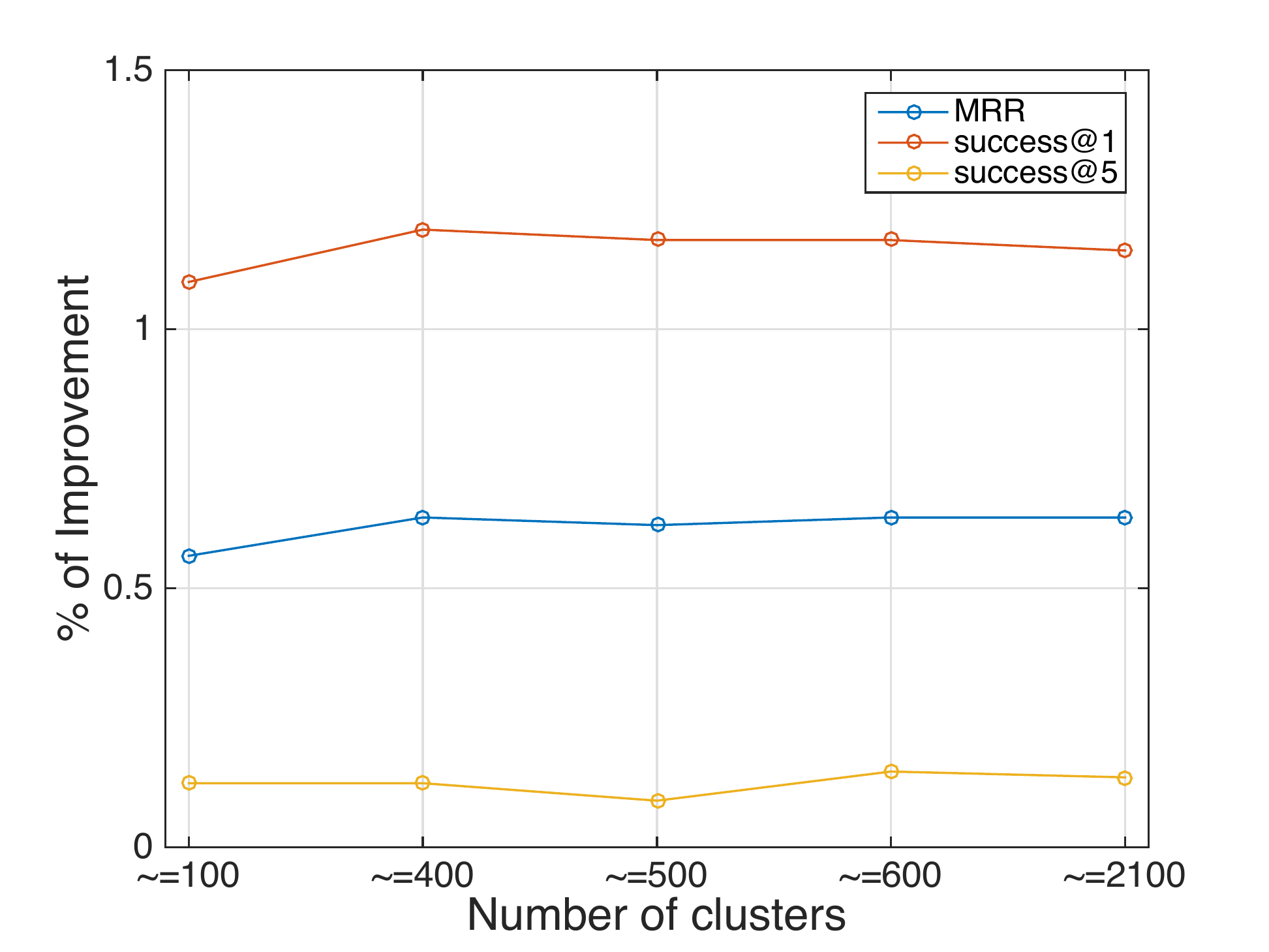}
    }
    \vspace{-0.3cm}
    \caption{(a) Most distinctive frequent $n$-grams for each query cluster. (b) Relative improvement of \qcwdnn~over \dnn~ v.s. \mixrate. (c) Relative improvement of \qcwdnn~over \dnn~v.s. cluster number.}
    \label{fig:analysis}
    \vspace{-0.3cm}
  \end{figure*}

\vspace{1mm}
\textbf{Query $n$-gram clusters.}
Due to the private nature of email search queries, we cannot show the query clustering results directly. Instead, we show the most distinctive $n$-gram features in Figure \ref{fig:query_cluster_features}.
The distinctiveness of each $n$-gram $s$ in cluster $c$ is defined as $\frac{cnt(s|c)}{cnt(s)}$, where $cnt(s)$ is the occurrence count of $s$ in the entire query set, and $cnt(s|c)$ is the occurrence count of $s$ in the queries of cluster $c$. 
These distinctive $n$-grams give us a hint about a query cluster's topic. 
As we can see, the queries about topic \emph{travel} are clustered together at the first level, which then be divided further into subtopics including \emph{car rental}, \emph{air flight}, and \emph{ride sharing}. 

\vspace{1mm}
\textbf{Effect of query cluster as raw feature.} 
To understand how the query cluster may contribute to the final ranking model as a raw input feature, we first feed it into the baseline \dnn~model as an additional query feature. Since there are hundreds and thousands of query clusters but each query can be assigned to at most ``number of tree depth'' clusters, we treat query cluster as a sparse feature and employ an embedding layer on top of it. Similarly, we generate cross-product features involving query cluster and feed them into the wide part of \qcwdnn.

\small
\begin{table}[!tbp]
\centering
\caption{\small Ranking performance of each proposed query-dependent ranking model. Relative performances compared with the baseline \dnn~model are shown in parentheses. The superscript $^{*}$ means the improvement is statistically significant.}
\vspace{-0.3cm}
\scalebox{0.86}{
\begin{tabular}{|l|l|l|l|}
\hline
\textbf{Method} & \textbf{MRR} & \textbf{success@1} & \textbf{success@5} \\
\hline
\dnn & 0.6698 & 0.4874 & 0.8861 \\
\qcdnn & 0.6697 (-0.01\%) & 0.4873 (-0.02\%) & 0.8864 (+0.03\%) \\
\qcwdnn & 0.6699 (+0.01\%) & 0.4875 (+0.02\%) & 0.8862 (+0.01\%) \\
\textbf{\qcmtlnn} & \textbf{0.6748 (+0.70\%)$^{*}$} & \textbf{0.4939 (+1.32\%)$^{*}$} & \textbf{0.8875 (+0.17\%)$^{*}$} \\
\hline
\end{tabular}
}
\vspace{-0.3cm}
\label{tab:singleMethod}
\end{table}
\normalsize

Table \ref{tab:singleMethod} reports the relative improvements achieved by the above two query cluster enhanced models. We can see that simply adding query cluster as a raw feature cannot improve the baseline \dnn~model. The reason is that the \dnn~architecture cannot differentiate between the query cluster feature and some other sparse query features. As for the \qcwdnn~model, it does allow query cluster features to contribute differently compared to other sparse query features via the feature crosses in the wide part. However, the query cluster is still treated as an input-level feature in the \qcwdnn~model.

\vspace{1mm}
\textbf{Effect of query cluster as separate label.} 
To reflect the intuition that query cluster information should guide how other features contribute to the ranking, we proposed the \qcmtlnn~model. Instead of considering the query cluster as a feature, the \qcmtlnn~model treats the query clusters as a separate ``label'' and learns how other query/document features can predict such a label. 
Results in Table \ref{tab:singleMethod} show that the \qcmtlnn~model significantly outperforms the already-strong baseline \dnn~model and can achieve 0.7\% and 1.32\% improvements in terms of MRR and success@1, respectively. 
As we evaluate our models on tens of millions of search queries, although the numbers may appear small, the improvements in practice are quite substantial. 
In fact, a +1\%~MRR improvement would be considered as a highly significant launch. 

\vspace{1mm}
\textbf{Effect of \mixrate.} 
During the training, \qcmtlnn~optimizes a joint loss including the log-loss of document preference prediction and the cross-entropy loss of query cluster prediction. 
To balance these two losses, we introduce a hyper-parameter \mixrate~in Eq. (6). 
In all previous experiments, we tune this hyper-parameter using a development set and then test the performance of \qcmtlnn~with this \mixrate fixed. 
As a record, the optimal \mixrate~tuned on development set is 0.9.
In this experiment, however, we intend to study how this hyper-parameter will influence the ranking performance and how \qcmtlnn~is sensitive to the choice of \mixrate.
Therefore, we train multiple \qcmtlnn~models with different \mixrate and directly report their performance on the testing data. 
The results are shown in Figure~\ref{fig:mixRateAnalysis}. First, we notice that a wide range of \mixrate choices can help improve \qcmtlnn's ranking performance (\emph{i.e.}, give us positive relative improvements). Second, we find that the performance of \qcmtlnn~first increases as \mixrate increases until it reaches about 1.8 and then starts to decrease when we further increase \mixrate. 
Notice that even the previous optimal \mixrate~tuned on development set is different from the optimal value 1.8 obtained directly on test set, our \qcmtlnn~model can still outperform baseline \dnn~model. 
We hypothesize that if we further increase the size of development set which gives more reliable estimation of \mixrate, the relative improvement of \qcmtlnn~model over \dnn~model will be more significant. 

%% Figure: 3 figures together

\vspace{1mm}
\textbf{Effect of query cluster number.} 
One important factor of \qcmtlnn~is the number of query clusters. To have an insight into how this factor may influence the ranking performance, we train multiple \qcmtlnn~models by varying the query cluster number and fixing all the other hyper-parameters. The result is plotted in Figure \ref{fig:clusterNumberAnalysis}. As we can see, the performance boost appears to not be very sensitive to the number of query clusters. We suspect the reason for this phenomenon is that around 100 query clusters have already covered most of the important data-dependent information.

%% End-to-end analysis
\vspace{-0.1cm}
\subsection{End-to-End Ranking Model Evaluation}
Nowadays, the production-level search engines commonly learn a final ranking function by leveraging multiple signals, which is also widely adopted in personal search scenario \cite{Wang2016LearningTR, Bendersky2017LearningFU, Zamani2017SituationalCF}. Therefore, in this experiment, we study whether we can use the output of \qcmtlnn~model (as a separate feature) alongside many other ranking signals to learn a global ranking function and improve the final performance.

\subsubsection{Experimental Setup}
\hfill

Following \cite{Zamani2017SituationalCF}, we use an adaptive learning-to-rank framework based on multiple adaptive regression tree (MART) algorithm \cite{Friedman1999GreedyFA}. 
Again, we apply the position bias correction techniques \cite{Wang2016LearningTR, Craswell2008AnEC} during the model training.

\subsubsection{Evaluation Metric}
\hfill

To evaluate the results from online production system, we use weighted mean reciprocal rank (WMRR) proposed in \cite{Wang2016LearningTR} and weighted average click position (WACP) as our primary metrics. 
They are calculated as follows:
\small
\begin{eqnarray*}
WMRR & = & \frac{1}{\sum_{i=1}^{|Q|} w_{i}} \sum_{i=1}^{|Q|} w_{i}  \cdot \frac{1}{rank_{i}}, \\
WACP & = & \frac{1}{\sum_{i=1}^{|Q|} w_{i}} \sum_{i=1}^{|Q|} w_{i} \cdot rank_{i},
\end{eqnarray*}
\normalsize
where $w_{i}$ denotes the bias correction weight, and it is inversely proportional to the probability of observing a click at the clicked position. We set those weights using result randomization, as described in \cite{Wang2016LearningTR}. Here, the lower WACP number indicates the better model performance\footnote{As the lower WACP number indicates the better model performance, a model reduces 4\% WACP is better than another one which reduces 3\% WACP.}.

\subsubsection{Results and Discussion}
\hfill

\begin{table}[!t]
\scriptsize
\centering
\caption{\small Relative improvements achieved by adding the output of each proposed model as a separate signal to our current personal ranking model, compared with the performance of the ranking model with only \dnn~model output as signal (denoted as LTR). The superscript $*$ means the improvement is statistically significant over the LTR and the superscript $**$ indicates the improvement is statistically significant over both the LTR and the LTR+\dnn.}.  
\vspace{-0.3cm}
\begin{tabular}{|l|l|l|}
\hline
\textbf{Method} & \textbf{WMRR} & \textbf{WACP} \\
\hline
LTR + \dnn & +2.35\%$^{*}$ & -3.24\%$^{*}$ \\
LTR + \qcdnn & +2.32\%$^{*}$ & -3.20\%$^{*}$ \\
LTR + \qcwdnn & +2.35\%$^{*}$ & -3.28\%$^{*}$ \\
\textbf{LTR + \qcmtlnn} & \textbf{+2.52\%$^{**}$} & \textbf{-3.41\%$^{**}$} \\
\hline
\end{tabular}
\label{tab:overAllMethod}
\vspace{-0.3cm}
\end{table}
In this experiment, we first adopt a general learning to rank (LTR) framework with many standard email search signals \cite{Carmel2015RankBT} as well as situational context signals \cite{Zamani2017SituationalCF}. We then feed the output of each of the \dnn~with/without query cluster, the \qcwdnn~with query cluster, and the \qcmtlnn~with query cluster as an additional feature into the learning to rank framework. In such an end-to-end setting, we can understand how much value will be added to the final ranking model. 

We show the relative improvement achieved by each model in Table \ref{tab:overAllMethod}. First, we can see that adding the \dnn~feature can significantly improve the overall ranking performance which confirms the previous finding \cite{Zamani2017SituationalCF}. Then, we notice that the performance boost by introducing \qcdnn~is less than simply using the raw \dnn~feature without query cluster, which may be caused by the noise in query clusters. Finally, we find that adding \qcmtlnn~feature can achieve the best performance boost among all the considered models.

%!TEX root = main.tex
% UTF-8 encoding
\section{Conclusions and Future Work}
In this paper, we study how to improve email search ranking by capturing query type information in an unsupervised fashion and constructing query-specific ranking models.
We first develop a hierarchical clustering algorithm based on truncated SVD and varimax rotation to obtain query type information. 
Then, we propose three query-dependent neural ranking models. 
The first two models leverage query type information as additional features, while the third one views query cluster as additional label and conducts document ranking and query cluster prediction simultaneously using multi-task learning. 
This multi-task learning scheme enables query type information to affect all intermediate layers of the neural model and guides the model to learn better query/document representations.
We evaluate our proposed approach on over 75 million real-world email search queries. 
The experiments demonstrate the effectiveness of our query clustering algorithm and show that our novel multi-task model can significantly outperform the baseline models which either do not incorporate query type information or just simply treat the query type as an additional feature. 

In the future, we would like to further study the following directions: 
(1) We may apply the current hierarchical clustering method to cluster users and similarly build a user-specific multi-task ranking model for email search. 
(2) We can extend the pairwise ranking paradigm to listwise ranking paradigm and it would be interesting to see how the query type information would help in listwise ranking models.
(3) We will explore how to choose the number and granularities of clusters that are most suitable for learning query-dependent ranking models.

\bibliographystyle{ACM-Reference-Format}
\vspace{-0.1cm}
\bibliography{sigproc}

\end{document}